\documentclass[aps,pre,amsmath,amsfonts,amssymb,floatfix,superscriptaddress,nofootinbib]{revtex4}
\usepackage{graphicx}  
\usepackage{dcolumn}   
\usepackage{bm}        
\usepackage{amssymb}   
\usepackage{amsmath}   

\usepackage{color}

\newcommand\be{\begin{equation}}
\newcommand\bea{\begin{eqnarray} \nonumber }
\newcommand\ee{\end{equation}}
\newcommand\eea{\end{eqnarray}}

\begin{document}

\title{The square-root impact law also holds for option markets}

\author{B.~T\'oth} \affiliation{Capital Fund Management, 23-25 Rue de l'Universit\'e, 75007 Paris, France.}
\author{Z.~Eisler} \affiliation{Capital Fund Management, 23-25 Rue de l'Universit\'e, 75007 Paris, France.}
\author{J.-P.~Bouchaud} \affiliation{Capital Fund Management, 23-25 Rue de l'Universit\'e, 75007 Paris, France.}\affiliation{CFM-Imperial Institute of Quantitative Finance, Department of Mathematics, Imperial College, 
180 Queen's Gate, London SW7 2RH}
\date{\today}

\begin{abstract}
Many independent studies on stocks and futures contracts have established
that market impact is proportional to the square-root of the executed
volume. Is market impact quantitatively similar for option markets as well?
In order to answer this question, we have analyzed the impact of a large
proprietary data set of option trades. We find that the square-root law indeed
holds in that case. This finding supports the argument for a universal
underlying mechanism.
\end{abstract}

\maketitle

The empirical study of market impact (i.e., the way trading influences prices in financial markets) is currently a hot topic in the field of market microstructure. 
Whereas one could have naively expected that trading a small quantity $Q$ moves on average the price proportionally to $Q$ \cite{Kyle}, many independent empirical studies have
demonstrated that the price change induced by the sequential execution of a total volume $Q$ follows an approximate $\sqrt{Q}$ law  
\cite{Barra:1997,GrinoldKahn,Almgren:2005,Moro:2009,Toth:2011,Iacopo:2013,Gomes:2013,Bershova2013,Brokmann:2014} (see however Ref. \cite{Farmer:new}). This square-root law is surprisingly \emph{universal}, independently of 
the type of contract traded (futures contracts, stocks or even Bitcoin \cite{Donier_bitcoin}), geographical zone, time period ($1995 \to 2014$), etc. For example the rise of high-frequency trading (HFT) in the last ten years seems 
to have had no effect on its validity (compare Refs.~\cite{Barra:1997,Almgren:2005} that use pre-2004 data with Refs.~\cite{Toth:2011,Iacopo:2013,Gomes:2013} that use post-2007 data). 

Motivated by this empirical observation, the idea of a ``latent order book'' \cite{Toth:2011} that has a locally linear V-shape around the current price has progressively 
gained momentum.  In particular, it was recently established in Ref.~\cite{Iacopo:PRL} that such a square-root impact indeed holds exactly within so-called ``reaction-diffusion'' 
models, where two types of particles (called $B$ and $A$), representing in a financial context the intended orders to buy (\emph{bid}) and to sell 
(\emph{ask}) diffuse on a line and disappear whenever they meet, corresponding to a transaction. The boundary between the $B$-rich region and the $A$-rich region therefore corresponds to the price $p_t$ \cite{Iacopo:PRL,J3}. 
This highly stylized model can be argued to capture the essential features of the liquidity
dynamics in financial markets, at least on long enough time scales where the precise microscopic details of the model become irrelevant (see Refs.~\cite{Iacopo:PRL,J3,Donier_W} for more insights). In fact, the mechanism leading to a V-shaped liquidity is so robust that one expects to observe an approximate square-root law in a very wide range of situations -- as alluded to above. One type of market where price impact has not 
-- to the best of our knowledge -- been empirically documented is that of options, where an {\it implied volatility} is priced by the participants. In view of the importance of these markets in modern finance, 
the absence of any available quantitative study of impact in option markets in somewhat surprising. 

Options are quite a bit more complex than stocks or futures contracts. Indeed, option prices move for several reasons. One is because the underlying
contract on which the option is written itself moves. Second is because time to maturity is always reducing. Third is because the future realized volatility forecast by market participants 
(called the implied volatility) is changing. Since the first effect can be approximately neutralized by holding an appropriate amount of the underlying
(the so-called \emph{delta hedge}) and the second effect is deterministic, option markets can in fact be seen as \emph{volatility markets}, where supply of volatility (i.e., risk insurers) meets demand of volatility 
(i.e., agents offloading risk). One therefore expects that an increased demand for volatility pushes up the implied volatility, and vice-versa -- much as the increased demand for a stock pushes its price up. 
The empirical question we ask here is: by how much? In order to answer this, we have analyzed the impact of CFM's option trades. As we hoped, we find that the impact is again strongly concave, and quantitatively comparable with what is observed on all other ``standard'' markets, or the Bitcoin.  This adds to the list of markets where the square-root law holds, and supports the case for the universality of the underlying mechanism. 

In the case of stocks or futures contracts, a {\it metaorder} corresponds to a certain quantity $Q$ of contracts that the manager (or the trading 
algorithm) decides to buy or sell. This quantity is then sent at time $t=0$ to the executing broker (or the execution algorithm) where it is sliced and diced and executed sequentially on markets until completion at time $t=T$. The average impact $I(Q)$ of the metaorder is defined as the average price change between $t=0^-$ and $t=T^+$. The square-root impact law discussed above states that:
\be
I(Q)=Y\cdot\sigma\cdot\sqrt{\frac{Q}{V}}, 
\label{sqrt_imp}
\ee
where $\sigma$ is the daily volatility of the asset, $V$ the daily traded volume and $Y$ a numerical constant of order unity (see e.g. Ref. \cite{Toth:2011}). Alternatively, one can write the difference between the total price paid during the metaorder and the price of the first trade (usually called implementation shortfall, or slippage) as:
\be
S(Q)=\int_0^Q I(q)dq = \frac23\cdot Y\cdot\sigma\cdot\sqrt{\frac{Q}{V}}\cdot Q.
\label{sqrt_slip}
\ee

For option markets, our definition of $Q$ is amended to take into account the fact that the same implied volatility can be traded using many different options, with different maturities and strikes. We will denote the net amount of \emph{vega} traded by CFM on a given day by $Q_\nu$, and the total gross amount of vega traded by the market by $V_\nu$. These correspond to the amount of directional volatility risk traded on a given underlying contract, after being delta hedged. We also define $\sigma_\sigma$, the so-called \emph{volatility-of-volatility}, as the standard deviation of the daily returns of the implied volatility of the options we traded, weighted by the traded vega amount. If the square-root impact law defined above for standard assets can be extended to option markets, one should expect the implementation shortfall measured after trading a size $Q_\nu$ to have the following form:
\be
S(Q_\nu)=\frac23\cdot  Y_\mathrm{vol}\cdot\sigma_\sigma\cdot\sqrt{\frac{Q_\nu}{V_\nu}}\cdot Q_\nu.
\label{sqrt_slip_opt}
\ee

\begin{figure}[t!]
  \centering 
  \includegraphics[scale=0.4]{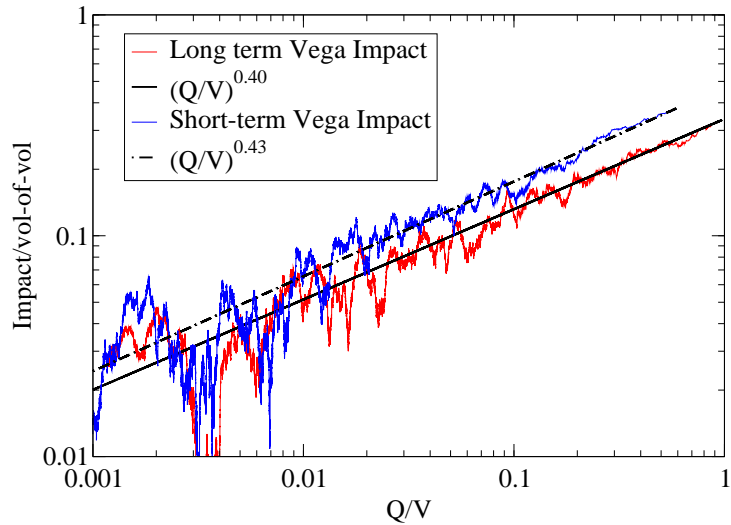}
 \caption{Running averages of the scatter plot of the rescaled impact $s=S/(Q_\nu\sigma_\sigma)$ \textit{vs}. the volume fraction $\phi := Q_\nu/V_\nu$ for 
 LT and ST options. The data has been shifted vertically to remove the linear cost term $b$ in Eq. (\ref{fit}) and allow a log-log representation that 
 illustrates the quality of the power-law fit. The fitted slopes, respectively $\delta_{LT} \approx 0.40$ and $\delta_{ST} \approx 0.43$, are close to the expected 
 universal $1/2$ value corresponding to the square-root impact law.}
\label{RegrFut}
\end{figure}

This is what we attempt to test below using our proprietary data set, which consists of 450,000 metaorders traded across options on more than 1000 single US stocks, in the period August 2013 to January 2016. Liquid options are on average mildly out-of-the-money (with a large dispersion) and have maturities ranging from a few days to over a year. We somewhat arbitrarily distinguish between short term (ST) options, with maturity $\leq 3$ months at the time of execution, and long term (LT) with maturity $> 3$ months at the time of execution, such that roughly half of the 450,000 metaorders are ST, the other half are LT. We measure implementation shortfall as the quantity weighted difference between the price of the traded options at the moment of the trade and their price at the moment when the metaorder decision was taken. We then rescale the instantaneous shortfall of each metaorder by the traded quantity $Q_\nu$ and the volatility-of-volatility of the considered options, and finally make a scatter plot of the result as a function of the volume fraction $\phi := Q_\nu/V_\nu$. For readability, we show in Fig. 1 the running average over 5,000 successive points of $s:=S/(Q_\nu\sigma_\sigma)$, and compare it to the best power-law fit of the form\footnote{The fits are made on both whole sets of $\sim 225,000$ points, after removing 
$\sim 50$ outliers for each set, and before doing the running average. Note that this running average is much smaller than the noise around it, which is by definition of order $1$ since the y-axis is divided by $\sigma_\sigma$.} 
\be\label{fit}
s = a \, \phi^\delta + b.
\ee
$\delta$ is the impact exponent, expected to be around $1/2$ and $b$ is the intercept that accounts for spread costs and/or gains from our execution 
signals allowing us to find optimal conditions for execution. Fig. 1 shows that these power-law fits are quite adequate in the range $\phi = 10^{-3} - 1$, and furthermore lead to a remarkably consistent value of $\delta_{LT} \approx 0.40$ and $\delta_{ST} \approx 0.43$. The values of $a$ and $b$ are found to be:
\be
a_{LT} \approx 0.33; \qquad b_{LT} \approx -0.013; \qquad a_{ST} \approx 0.40; \qquad b_{ST} \approx -0.071.
\ee
We note that the $Y$ constant obtained for stocks and futures contracts is in the range $0.5 \to 1$. Remarkably, we find that $Y_{\text{vol}} = 3a/2 \sim 0.5 - 0.6$ is exactly in the 
same range. The impact of trades on the implied volatility of single stock options is therefore fully compatible with all the results established so far on 
other contracts. We have also studied the impact on the implied volatility of futures contracts (stock indices, commodities, bonds, currencies) but our data set only contains $10,000$ metaorders. Although much more noisy, the results are again compatible with a concave impact law.

The conclusion of this short note is that, as announced in the title, the square-root impact law also holds for option markets, provided one defines 
the size of the metaorder as the net vega. The resulting impact law is quantitatively similar to what is observed in all other markets where it 
has been documented, including the overall dimensionless constant $Y_{\text{vol}}$ that appears in Eq. (\ref{sqrt_slip_opt}). The empirical study presented here is,
to our knowledge, the first quantitative result on market impact for options, and should be relevant for option traders. The fact that impact in option markets is fully 
compatible with all previously published results \cite{Barra:1997,GrinoldKahn,Almgren:2005,Moro:2009,Toth:2011,Iacopo:2013,Gomes:2013,Bershova2013,Brokmann:2014}, including the Bitcoin \cite{Donier_bitcoin}, 
bolsters the hypothesis that the square-root impact is universal across all traded markets, 
and vindicates the idea of a locally linear latent liquidity profile around the traded price, advocated in Refs.~\cite{Toth:2011,Iacopo:PRL,J3,Donier_W}.

\vskip 0.5cm

We thank J. Donier, I. Mastromatteo, C. Lehalle, J. Kockelkoren and M. Potters for many inspiring discussions.

\end{document}